\documentclass[twocolumn,prc,aps,floatfix]{revtex4}
\usepackage[dvips]{graphicx}
\begin{document}
\title{Limits on temporal  variation of quark masses and strong
interaction from atomic clock experiments}
\author{V.V. Flambaum}
\affiliation{
 School of Physics, The University of New South Wales, Sydney NSW
2052, Australia
}
\date{\today}
\begin{abstract}
We perform calculations of the dependence of nuclear magnetic moments
on quark masses and
obtain limits on the variation of
$(m_q/\Lambda_{QCD})$
 from recent atomic clock
experiments with hyperfine transitions in H, Rb, Cs, Hg$^+$
and optical transtion in Hg$^+$.
\end{abstract}
\maketitle


\section{Introduction}
 Interest in the temporal and spatial variation of  major constants of physics
has been recently revived by
 astronomical data which seem to suggest a variation
 of the electromagnetic constant
 $\alpha=e^2/\hbar c$   at the $10^{-5}$ level
 for the time scale 10 billion years, see \cite{alpha}
 (a discussion of
other limits can be found in the review \cite{uzan} and references therein).
 However, an independent experimental confirmation is needed.

The hypothetical unification of all interactions implies that variation
of the electromagnetic interaction constant $\alpha$ should be accompanied
by the variation of masses and the strong interaction constant.
 Specific predictions need a model. For example, the grand unification
model  discussed in \cite{Langacker:2001td} predicts that
 the
quantum chromodynamic (QCD) scale  $\Lambda_{QCD}$
 (defined as the position of the Landau pole in the logarithm for the
running strong coupling constant)
is modified as follows
\begin{equation} \label{QCD}
{\delta \Lambda_{QCD} \over  \Lambda_{QCD}}\approx 34 {\delta \alpha
\over \alpha}
\end{equation}
The variation of quark and electron masses in this model is  given by
\begin{equation} \label{mq}
{\delta m \over m}\sim 70 {\delta \alpha \over \alpha}
\end{equation}
This gives an estimate for the variation of the dimensionless ratio
\begin{equation} \label{mQCD}
{\delta(m/ \Lambda_{QCD}) \over(m/\Lambda_{QCD})}\sim 35 {\delta \alpha
\over \alpha}
\end{equation}
The large coefficients in these expressions are generic for grand unification
models, in which modifications come from high energy scales:
they appear because the running strong coupling constant and
 Higgs constants (related to mass) run faster than $\alpha$.
 This means that if these models
are correct the variation
of masses and strong interaction may be easier to detect than the variation
of $\alpha$.

 Unlike for the electroweak forces, for
the strong interaction there is generally no direct relation between
the coupling constants and observable quantities.
 Since one can measure only variation of the
dimensionless quantities, we want to extract from the measurements
 variation
of the dimensionless ratio $m_q/\Lambda_{QCD}$ where $m_q$ is the quark
mass (with the dependence on the normalization point removed).
A number of limits on variation of $m_q/\Lambda_{QCD}$
have been obtained recently from consideration
of Big Bang Nucleosynthesis, quasar absorption spectra
and Oklo natural nuclear reactor which was active about
1.8 billion years ago \cite{FS,oliv,dmitriev,FS1} (see also
\cite{Murphy1,Cowie,Oklo,c12,savage}). Below we consider
the limits which follow from laboratory atomic clock comparison.
Laboratory limits with a time base about a year are
especially sensitive to oscillatory variation of
fundamental constants. A  number of relevant measurements
have been performed already and even larger number  have been started
or planned. The increase in precision is very fast.

It has been pointed out by Karshenboim \cite{Karschenboim}
 that measurements  of ratio of hyperfine structure intervals
in different atoms are sensitive to
variation of nuclear magnetic moments.
 First rough estimates of nuclear magnetic moments dependence on
 $m_q/\Lambda_{QCD}$ and limits on time variation of this ratio
have been obtained in our paper
\cite{FS}. Using
H, Cs and Hg$^+$ measurements \cite{prestage,Cs},
 we obtained the limit on variation of $m_q/\Lambda_{QCD}$
 about $5 \cdot 10^{-13}$ per year.
Below we calculate the dependence of nuclear magnetic moments
 on  $m_q/\Lambda_{QCD}$ and  obtain the limits
 from recent atomic clock
experiments with hyperfine transitions in H, Rb, Cs, Hg$^+$
and optical transition in Hg$^+$.
It is convenient to assume that the strong interaction scale
 $\Lambda_{QCD}$ does not vary, so we will speak about variation
of masses.

     The hyperfine structure constant can be presented in the following
form
\begin{equation}\label{A}
A=const \times [\frac{m_e e^4}{\hbar ^2}] [ \alpha ^2 F_{rel}( Z \alpha)]
[\mu \frac{m_e}{M_p}]
\end{equation}
The factor in the first bracket is an atomic unit of energy. The second
``electromagnetic'' bracket determines the dependence on $\alpha$.
An approximate expression for the relativistic correction factor (Casimir
factor) for s-wave electron is the following
\begin{equation}\label{F}
F_{rel}= \frac{3}{\gamma (4 \gamma^2 -1)}
\end{equation}
where $\gamma=\sqrt{1-(Z \alpha)^2}$, Z is the nuclear charge.
Variation of $\alpha$ leads to the following variation of $F_{rel}$
 \cite{prestage}:
\begin{equation}\label{dF}
\frac{\delta F_{rel}}{F_{rel}}=K \frac{\delta \alpha}{\alpha}
\end{equation}
\begin{equation}\label{K}
K=\frac{(Z \alpha)^2 (12 \gamma^2 -1)}{\gamma^2 (4 \gamma^2 -1)}
\end{equation}
More accurate numerical many-body calculations \cite{dzuba1999}
 of the dependence of the hyperfine structure on $\alpha$ have shown
 that the coefficient $K$ is slightly larger than that given by this
formula. For Cs ($Z$=55) $K$= 0.83 (instead of 0.74),
for Rb $K$=0.34
(instead of 0.29),  for Hg$^+$
$K$=2.28 (instead of 2.18).

    The last bracket in eq. (\ref{A})  contains the dimensionless
 nuclear magnetic moment $\mu$ in nuclear magnetons
 ( the nuclear magnetic moment $M=\mu\frac{e\hbar}{2 M_p c}$),
 electron mass $m_e$ and proton mass $M_p$. We may also include
a small correction due to the finite nuclear
size. However, its contribution is insignificant.

Recent experiments measured time dependence of the ratios of
 hyperfine structure intervals of $^{199}$Hg$^+$ and H \cite{prestage},
$^{133}$Cs and $^{87}$Rb \cite{marion} and ratio of optical frequency
 in Hg$^+$ and $^{133}$Cs hyperfine frequency \cite{bize}.
 In the ratio of two
hyperfine structure constants for different atoms time dependence
may appear from the ratio of the factors $F_{rel}$ (depending on $\alpha$)
 and ratio of nuclear magnetic moments (depending on $m_q/\Lambda_{QCD}$).
Magnetic moments in a single-particle approximation (one unpaired nucleon)
 are:
\begin{equation}\label{mu+}
\mu=(g_s + (2 j-1) g_l)/2
\end{equation}
for $j=l+1/2$.
\begin{equation}\label{mu-}
\mu=\frac{j}{2(j+1)}(-g_s + (2 j+3) g_l)
\end{equation}
for $j=l-1/2$. Here the orbital g-factors are
 $g_l=1$ for valence proton and $g_l=0$ for valence
neutron. The present values of spin g-factors $g_s$ are
$g_p=5.586$ for proton and $g_n=-3.826$ for neutron.
 They depend on $m_q/\Lambda_{QCD}$.
The light quark masses are only about $1 \%$ of the nucleon mass
 ($m_q=(m_u+m_d)/2 \approx$ 5 MeV). The nucleon magnetic moment remains
 finite in the chiral  limit of $m_u=m_d=0$. Therefore, one may think that
 the corrections to $g_s$ due to the finite quark masses are very small.
However, there is a mechanism which enhances quark mass contribution:
$\pi$-meson loop corrections to the nucleon magnetic moments which
are proportional to $\pi$-meson mass $m_{\pi} \sim \sqrt{m_q\Lambda_{QCD}} $;
$m_{\pi}$=140 MeV is not so small.

    According to  calculation in Ref. \cite{Thomas} dependence of the
nucleon g-factors
on $\pi$-meson mass $m_\pi$ can be approximated by the following equation
\begin{equation}\label{thomas}
g(m_\pi)=\frac{g(0)}{1+ a m_\pi + b m_\pi ^2}
\end{equation}
where $a$= 1.37/GeV ,  $b$= 0.452/GeV$^2$ for proton
and  $a$= 1.85/GeV ,  $b$= 0.271/GeV$^2$ for neutron. This leads to
the following estimate:
\begin{equation}\label{gp}
\frac{\delta g_p}{g_p} =
 -0.174 \frac{\delta m_\pi}{m_\pi}= -0.087 \frac{\delta m_q}{m_q}
\end{equation}
\begin{equation}\label{gn}
\frac{\delta g_n}{g_n} =
 -0.213 \frac{\delta m_\pi}{m_\pi}= -0.107 \frac{\delta m_q}{m_q}
\end{equation}
Eqs. (\ref{mu+},\ref{mu-},\ref{gp},\ref{gn}) give variation
 of nuclear magnetic moments.
For hydrogen nucleus (proton)
\begin{equation}\label{H}
\frac{\delta \mu}{\mu} =\frac{\delta g_p}{g_p}= -0.087 \frac{\delta m_q}{m_q}.
\end{equation}
For $^{199}$Hg we have valence neutron (no orbital contribution),
therefore the result is
\begin{equation}\label{Hg}
\frac{\delta \mu}{\mu} =
\frac{\delta g_n}{g_n} = -0.107 \frac{\delta m_q}{m_q}
\end{equation}
For $^{133}$Cs we have valence proton with $j$=7/2, $l$=4 and
\begin{equation}\label{Cs}
\frac{\delta \mu}{\mu} =
 0.22 \frac{\delta m_\pi}{m_\pi}= 0.11 \frac{\delta m_q}{m_q}
\end{equation}
For $^{87}$Rb we have valence proton with $j$=3/2, $l$=1 and
\begin{equation}\label{Rb}
\frac{\delta \mu}{\mu} =
 -0.128 \frac{\delta m_\pi}{m_\pi}=- 0.064 \frac{\delta m_q}{m_q}
\end{equation}
Deviation of the single-particle values of nuclear magnetic moments from
the measured values is about 30 $\%$. Therefore, we tried
to refine the single-particle estimates. If we neglect spin-orbit
interaction the total spin of nucleons is conserved.
The magnetic moment of nucleus changes due to the spin-spin
interaction because valence proton
transfers a part of its spin $<s_z>$ to core neutrons (transfer of spin
from the valence proton to core protons does not change the magnetic moment).   In this approximation $g_s=(1-b)g_p + b g_n$ for valence proton
 (or $g_s=(1-b)g_n + b g_p$ for valence
 neutron). We can use coefficient
$b$ as a fitting parameter to reproduce nuclear magnetic moments
exactly. The sign of $g_p$ and $ g_n$ are opposite, therefore
a small mixing $b \sim 0.1$ is enough to eliminate the deviation
of the theoretical value from the experimental one.
Note also that it follows from eqs. (\ref{gp},  \ref{gn}) that
$\frac{\delta g_p}{g_p} \approx \frac{\delta g_n}{g_n}$.
This produces an additional suppression of the effect
of the mixing. This indicates that the actual accuracy
of the single-particle approximation for the effect of the spin
g-factor variation may be as good as 10 $\%$.
Note, however, that here we neglected variation of the
mixing parameter $b$ which is hard to estimate.

   Now we can estimate sensitivity of the ratio of the hyperfine
transition frequencies to variation of $m_q/\Lambda_{QCD}$.
For $^{199}$Hg and hydrogen we have
\begin{equation}\label{HgH}
\frac{\delta[A(Hg)/A(H)]}{[A(Hg)/A(H)]}=
2.3 \frac{\delta \alpha}{\alpha}
 -0.02 \frac{ \delta [m_q/\Lambda_{QCD}]}{[m_q/\Lambda_{QCD}]}
\end{equation}
Therefore, the measurement of the ratio of Hg and hydrogen
hyperfine frequencies is practically insensitive to the variation
of masses and strong interaction.
The result of measurement \cite{prestage}  may be presented
as a limit on variation of the parameter $\tilde{\alpha}=
\alpha [m_q/\Lambda_{QCD}]^{-0.01}$:
\begin{equation}\label{limitHgH}
 |\frac{1}{ \tilde{\alpha}}\frac{d \tilde{\alpha}}{dt}|
 < 3.6 \times 10^{-14}/year
\end{equation}
Other ratios of hyperfine frequencies are more sensitive to
$m_q/\Lambda_{QCD}$. For $^{133}$Cs/$^{87}$Rb we have
\begin{equation}\label{CsRb}
\frac{\delta[A(Cs)/A(Rb)]}{[A(Cs)/A(Rb)]}=
0.49 \frac{\delta \alpha}{\alpha}
 +0.17 \frac{ \delta [m_q/\Lambda_{QCD}]}{[m_q/\Lambda_{QCD}]}
\end{equation}
Therefore, the result of the measurement \cite{marion}  may be presented
as a limit on variation of the parameter $X=
\alpha^{0.49} [m_q/\Lambda_{QCD}]^{0.17}$:
\begin{equation}\label{limitCsRb}
 \frac{1}{X}\frac{dX}{dt}=
 (0.2 \pm 7) \times 10^{-16}/year
\end{equation}
Note that if the relation (\ref{mQCD}) is correct, variation of $X$
would be dominated by  variation of $[m_q/\Lambda_{QCD}]$.
 The relation (\ref{mQCD}) would give
$X \propto \alpha^{7}$ and limit on $\alpha$ variation
 $\frac{1}{\alpha}\frac{d\alpha}{dt}=
 (0.03 \pm 1) \times 10^{-16}/year$ .

For $^{133}$Cs/H we have
\begin{equation}\label{CsH}
\frac{\delta[A(Cs)/A(H)]}{[A(Cs)/A(H)]}=
0.83 \frac{\delta \alpha}{\alpha}
 +0.2 \frac{ \delta [m_q/\Lambda_{QCD}]}{[m_q/\Lambda_{QCD}]}
\end{equation}
Therefore, the result of the measurements \cite{Cs}  may be presented
as a limit on variation of the parameter $X_H=
\alpha^{0.83} [m_q/\Lambda_{QCD}]^{0.2}$:
\begin{equation}\label{limitCsH}
 |\frac{1}{X_H}\frac{dX_H}{dt}|<
 5.5 \times 10^{-14}/year
\end{equation}
If we assume the relation (\ref{mQCD}), we would have
$X_H \propto \alpha^{8}$, $|\frac{1}{\alpha}\frac{d\alpha}{dt}|<
0.7 \times 10^{-14}/year$.

 The  optical clock transition energy $E(Hg)$ ($\lambda$=282 nm)
in Hg$^+$ ion  can be presented in the following form:
\begin{equation}\label{E}
E(Hg)=const \times [\frac{m_e e^4}{\hbar ^2}] F_{rel}(Z \alpha)
\end{equation}
Note that the atomic unit of energy (first bracket) is canceled out in ratios, therefore,
we should not consider its variation. Numerical calculation
of  the relative variation
of $E(Hg)$ has given  \cite{dzuba1999}:
\begin{equation}\label{dE}
\frac{\delta E(Hg)}{E(Hg)}=-3.2\frac{\delta \alpha}{\alpha}
\end{equation}
Variation of the ratio of the Cs hyperfine splitting $A(Cs)$
to this optical transition energy is equal to
\begin{equation}\label{CsHgE}
\frac{\delta[A(Cs)/E(Hg)]}{[A(Cs)/E(Hg)]}=
6.0 \frac{\delta \alpha}{\alpha} +
 \frac{ \delta [m_e/\Lambda_{QCD}]}{[m_e/\Lambda_{QCD}]}
 +0.11 \frac{ \delta [m_q/\Lambda_{QCD}]}{[m_q/\Lambda_{QCD}]}
\end{equation}
Here we have taken into account that the proton mass
$M_p \propto \Lambda_{QCD}$. The factor 6.0 before $\delta \alpha$
 appeared from $\alpha^2 F_{rel}$
in the Cs hyperfine constant (2+0.83) and $\alpha$-dependence of $E(Hg)$
(3.2).  Therefore, the work \cite{bize}
gives  the limit on variation of the parameter $U=
\alpha^{6} [m_e/\Lambda_{QCD}][m_q/\Lambda_{QCD}]^{0.1}$:
\begin{equation}\label{limitCsHgE}
 |\frac{1}{U}\frac{dU}{dt}|<
 7 \times 10^{-15}/year
\end{equation}
If we assume the relation (\ref{mQCD}), we would have
$U \propto \alpha^{45}$, $|\frac{1}{\alpha}\frac{d\alpha}{dt}|<
1.5 \times 10^{-16}/year$. Note that we presented such limits
on $\frac{1}{\alpha}\frac{d\alpha}{dt}$ as an illustration only
since they are strongly model-dependent.

This work is  supported by the Australian Research
Council.

\end{document}